\documentclass[11pt,twoside]{article}

\usepackage[usenames,x11names,table]{xcolor}
\usepackage{float}
\usepackage[font={rm,md,sl},format=hang]{subfig}
\usepackage[font={rm,small,sl}]{caption}
\usepackage{graphicx}
\usepackage[obeyspaces]{url}% http://ctan.org/pkg/url
\usepackage[pagebackref,breaklinks]{hyperref}
\usepackage{breakurl}
\usepackage[margin=1.1in]{geometry}
\usepackage{enumerate}
\usepackage{tikz}
\usepackage{amsmath}

\usepackage{listings}

\hypersetup{%
	pdfusetitle=true,%
	bookmarks=true,%
	pdftitle={NoDBA: Automatic Database Administration using Deep Reinforcement Learning},%
    pdfauthor={Ankur Sharma, Felix Martin Schuhknecht, Jens Dittrich},     % author
    pdfsubject={Databases, Machine Learning},   % subject of the document
    pdfcreator={Ankur Sharma, Felix Martin Schuhknecht, Jens Dittrich},   % creator of the document
    pdfproducer={\LaTeX}, 
	pdfkeywords={Deep Learning, Reinforcement Learning, Machine Learning},%
	pdfdisplaydoctitle=true,%
	pdfstartview=Fit,%
	colorlinks=false%
}

\renewcommand*{\backrefalt}[4]{%
\ifcase #1 %
No citations.%
\or
One citation on page #2.%
\else
#3 citations on pages #2.%
\fi
}

\colorlet{color}{cyan!50}

\newcolumntype{L}[1]{>{\raggedright\let\newline\\\arraybackslash\hspace{0pt}}m{#1}}
\newcolumntype{C}[1]{>{\centering\let\newline\\\arraybackslash\hspace{0pt}}m{#1}}
\newcolumntype{R}[1]{>{\raggedleft\let\newline\\\arraybackslash\hspace{0pt}}m{#1}}

\begin{document}

% ****************** TITLE ****************************************

\title{The Case for Automatic Database Administration using\\Deep Reinforcement Learning}

% ****************** AUTHORS **************************************

\author{Ankur Sharma\qquad Felix Martin Schuhknecht\qquad Jens Dittrich\\
\\
Big Data Analytics Group\\
Saarland Informatics Campus\\
\url{https://infosys.uni-saarland.de}}

\date{\today}

\maketitle

\begin{abstract}

Like any large software system, a full-fledged DBMS offers an overwhelming amount of configuration knobs. These range from static initialisation parameters like buffer sizes, degree of concurrency, or level of replication to complex runtime decisions like creating a secondary index on a particular column or reorganising the physical layout of the store. To simplify the configuration, industry grade DBMSs are usually shipped with various advisory tools, that provide recommendations for given workloads and machines. However, reality shows that the actual configuration, tuning, and maintenance is usually still done by a human administrator, relying on intuition and experience.

Recent work on deep reinforcement learning has shown very promising results in solving problems, that require such a sense of intuition. For instance, it has been applied very successfully in learning how to play complicated games with enormous search spaces. Motivated by these achievements, in this work we explore how deep reinforcement learning can be used to administer a DBMS. First, we will describe how deep reinforcement learning can be used to automatically tune an arbitrary software system like a DBMS by defining a problem environment. 
Second, we showcase our concept of NoDBA at the concrete example of index selection and evaluate how well it recommends indexes for given workloads.

\end{abstract}

\section{Introduction}
\label{sec:introduction}
An industry-grade DBMS is a massive software system. As the requirements of the individual end-users and their workloads heavily differ, these systems are equipped with tons of options and configuration knobs. Finding the "right" setup for the current situation has been an extremely challenging task ever since and even lead to the creation of an entire profession around that task: the \textit{database administrator} or \textit{DBA}. These administrators constantly monitor the performance of the system and tune the parameters to fit to the current workload as good as possible. To a certain degree, they rely on so called \textit{design advisory tools}, that provide recommendations on what to do based on statistics. However, as statistics have limits~\cite{LeisGMBK015}, a good administrator will always rely at least equally on experience and intuition to come to the right decision.

Both \textit{experience} and \textit{intuition} are deeply embedded in the field of machine learning. Recently, deep learning techniques showed fascinating results in learning complex tasks, such as playing difficult games~\cite{alphago, ataripaper}. In general, they seem to perform great in situations with vast search spaces and problems, that are too complex to grasp by traditional machine learning approaches~\cite{alphago,dittrich17}.
This raises the question whether it is possible to apply deep learning techniques to the problem of DBMS administration as well. Due to the vast amount of configuration parameters and workload differences, the search space is huge as well and extremely hard to overview --- the perfect candidate for deep learning.

\subsection{Deep Reinforcement Learning}
\label{ssec:drl}
Especially interesting in the context of administration is \textit{deep reinforcement learning}~\cite{drl}. In contrast to traditional supervised learning~\cite{supervised}, where a neural network is trained on a set of given inputs and expected outputs, in reinforcement learning, the training process does not require any expected outputs. Instead, the training is completely driven by so called~\textit{rewards}, that tell the learner whether a taken action lead to a positive or a negative result on the input. Depending on the outcome, the neural network is encouraged or discouraged to consider the action on this input in the future. Obviously, this is directly equivalent to an administrator, who changes a parameter of the system setup and is either rewarded or punished with an improved or worsened performance. A smart DBA will base future decisions on the made experiences. 

In the same manner, we can apply deep reinforcement learning to train a neural network in taking over the administration process. Essentially, we have to define a so called problem environment consisting of the four following components to perform the learning:

\begin{enumerate}[(1)]

\item \textbf{The \textit{input} to the neural network.} This is typically the current workload in form of query characteristics, for which the system should be optimised as well as the current state of the configuration.

\item \textbf{The \textit{set of actions}, that can be taken.} An action could be to create a secondary index on a certain column or to change the size of a database buffer. Such an action is a transition from a current system configuration to a new system configuration. 

\item \textbf{The \textit{reward function}, that rates the impact of a taken action.} For example, the runtime of the new configuration after the action has been taken could be compared with the best configuration seen so far. The higher the improvement, the higher is the returned positive reward. If performance degrades, a negative reward is returned.
 
\item \textbf{The \textit{hyper parameters} to steer the learning process.} This includes properties of the neural network (e.g. number of hidden layers, number of nodes per layer) as well as properties of the learning process like the number of iterations. 

\end{enumerate}

With the right instantiation of these $4$~categories, we are generally able to train a neural network for a given optimisation goal (e.g. minimal runtime) and a given workload (e.g. a set of queries).

\subsection{Prediction and Training}
\label{ssec:learner}

With a high-level understanding of the involved components, let us now first go through the workflow of a general learner and see how it predicts and trains. Assuming that all required hyper parameters are configured, which we will discuss in detail in Section~\ref{ssec:parameters}, step~$i$ of the learning process essentially consists of the following sequence:

\begin{enumerate}
\item Using the neural network, we predict the next action~$a$ to perform. In the beginning of the training phase, a prediction will be mostly arbitrary and of low quality. Over time, the training will (hopefully) improve the prediction quality. 
 
\item We apply the predicted action~$a$ to the current configuration~$L_{i-1}$, resulting in a new configuration~$L_{i}$.
\item We compute the reward~$r$ of the taken action~$a$ using the reward function on the new configuration. This transition from one configuration to a new configuration can be summarised in the quadruple $$Q_i=(L_{i-1}, a, L_{i}, r)$$

\item We train the neural network using $Q_i$ and restart in (1). A positive reward~$r$ will enforce the neural network to predict~$a$ on future configurations, which are similar to~$L_{i-1}$. 

\end{enumerate}

\begin{figure}[!htb]

	\begin{center}
		\includegraphics[width=7cm]{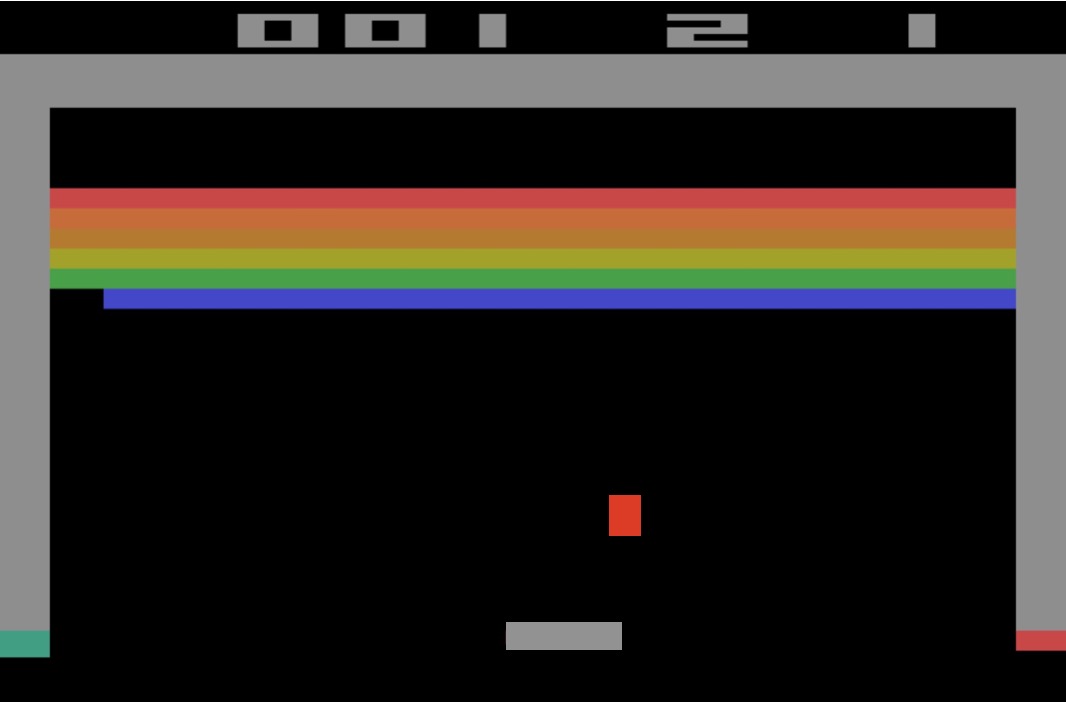}
	\end{center}
	\vspace{-0.3cm}
	\caption{Breakout on the Atari 2600}
	\label{figs:breakout}
\end{figure}

For a better understanding, let us see how this general workflow maps to the concrete example of playing a famous classic Atari game: Breakout~\cite{ataripaper}.
The goal of Breakout is to destroy a set of blocks, located at the top of the screen, by repeatedly bouncing back a ball with a paddle, that can be moved horizontally. As soon as all blocks are destroyed, the game is won. If the paddle misses the ball, such that the ball drops below the screen, the game is lost. Figure~\ref{figs:breakout} shows Breakout on the Atari 2600. 

To perform reinforcement learning on Breakout, the current state of the game board (i.e. the locations of blocks, ball, and paddle) is fed into the neural network as input. The set of actions is composed of the possible movements of the paddle: the player can either move a step to the left, move a step to the right, or stay. A positive reward is generated, if the paddle hits the ball. Table~\ref{table:game_vs_index_selection} summarises the setup again.

\begin{table}[!htb]
	\small
	\setlength{\tabcolsep}{3pt}
	\begin{center}
		\begin{tabular}{| L{2cm} || R{6.6cm} | R{6.6cm} |}
		\hline
			\textbf{Category} & \textbf{Atari Breakout~\cite{ataripaper}} & \textbf{NoDBA [this work]}\\
			\hline
			% 50 cols, small pages
			%\hline
			\textbf{Input} & Current state of the game board. & Workload and current index configuration.\\
			\hline
			 \textbf{Set of Actions} & Move paddle left, move paddle right, stay & Create an index on a particular column. \\
			\hline
			\textbf{Reward} & Positive, if the paddle hits the ball. & Positive, if an index configuration improves over previous state.\\
			\hline
			%\textbf{Visual Concept} & \includegraphics[width=.4\columnwidth]{breakout.jpg} & \includegraphics[width=.4\columnwidth]{nodba.jpg}\\
			%\hline
		\end{tabular}
		\caption{Comparison of Input, Set of Actions, and Reward for Atari Breakout and NoDBA.}
		\label{table:game_vs_index_selection}
	\end{center}
\end{table}

The actual learning happens by applying the previously described steps. Early in the learning phase, the predicted paddle movements seem random and meaningless. However, as soon as the paddle manages to bounce back the ball, a positive reward is generated and fed back into the network for training. Over time, the system will learn to position the paddle horizontally as close as possible to the falling ball.      

Obviously, the core principle of deep reinforcement learning is fairly simple and thanks to the general concepts of input, action set, and reward function, can be applied to a variety of problems.  
In this paper, we will showcase the quality of deep reinforcement learning in the database world using \textit{index selection} as an example --- a task that has a crucial impact on query execution times.

\subsection{Automatic Index Selection}
\label{ssec:indexsel}
Given a workload, the task of index selection is to decide on which attributes to create secondary indexes, such that the processing of the workload benefits the most. Typically, an upper limit on the number of indexes to create is given due to space constraints and index maintenance costs, so simply indexing all columns is not an option. Of course, the goal is to come up with a proper index selection in an automatic way without the need of a human administrator. Let us precisely define the problem: assuming our database schema~$S$ consists of $n$~columns (which can originate from multiple tables) and $k$~is the maximum amount of indexes we are allowed to create with $k \leq n$, then we want to find a subset of $S$ of maximum size~$k$ that minimises the runtime on a workload~$W$.

\section{NoDBA}
With the problem definition in mind, let us see step by step how we can map the problem of index selection to the general deep reinforcement learner described previously in Section~\ref{ssec:learner}. We will first define the input to the neural network in Section~\ref{ssec:input}, the set of possible actions in Section~\ref{ssec:actions}, and the reward function in Section~\ref{ssec:reward}. Then, we will discuss the hyper parameters of the learner in Section~\ref{ssec:parameters}, that we have to set up. 

\subsection{Neural Network Input} 
\label{ssec:input}

We start by defining the input to our neural network. As mentioned in Section~\ref{ssec:drl}, this input is typically a combination of the encoding of (1)~the workload and (2)~the current configuration. Thus, we will separate the input into two parts~$I_{workload}$ and $I_{indexes}$, that are both fed into the neural network. 

In $I_{workload}$, we encode the characteristics of the workload in form of a matrix of size~$n \cdot m$, where $n$ is the number of queries in the workload and $m$~is the number of columns in the database schema.
An entry~$i,j$~($i<n, j<m$) of the matrix describes for a query~$Q_i$ and a column~$C_j$ the selectivity~$Sel(Q_i,C_j)$:
$$I_{workload}=\begin{bmatrix}
    Sel(Q_0,C_0)       & \dots & Sel(Q_0,C_{m-1}) \\
    \vdots       & \ddots & \vdots \\
    Sel(Q_{n-1},C_0)       & \dots & Sel(Q_{n-1},C_{m-1}) 
\end{bmatrix}$$
with
\begin{equation*}
Sel(Q_i,C_j) = 
\begin{cases}
\frac{\text{\# selected records of}~Q_i~\text{on}~C_j}{\text{\# total records of}~C_j}  & \small{\text{if}~Q_i~\text{selects on}~C_j} \\
1 & \small{\text{if}~Q_i~\text{does not select on}~C_j}
\end{cases}
\end{equation*}

If a query $Q_i$ selects on a particular column~$C_j$, then $Sel(Q_i, C_j)$ returns the number of selected records in relation to the total number of records --- the smaller the result, the higher\footnote{A \textit{small} number of returned records is called a \textit{high} selectivity.} the selectivity. If a query~$Q_i$ does not select on a particular column~$C_j$ at all, then~$Sel(Q_i, C_j)=1$. This denotes the lowest possible selectivity and will encourage the system \textit{not} to consider building an index on this column. 
These selectivities are computed by running every query of the workload on the database once upfront. 

As already mentioned, we additionally have to encode the current index configuration in~$I_{indexes}$, which describes on which columns indexes exist. Thus, $I_{indexes}$ is simply the following bitlist of size~$m$
$$I_{indexes}=\begin{bmatrix}
    hasIndex(C_0)       & \dots & hasIndex(C_{m-1})
\end{bmatrix}$$
with $hasIndex(C_j)=1$, if there exists an index on $C_j$ or no query in the workload uses~$C_j$, and $0$~otherwise.

\subsection{Set of Actions}
\label{ssec:actions}
Next, we have to define the set of possible actions~$A$, that can be carried out. For the problem of index selection, we have to consider only a single action: the creation of an index on a specific column\footnote{We do not consider multi-column indexes in this use-case.}~$C_j$. Thus, $$A=\{\texttt{create\_index\_on(}C_j\texttt{)}\}$$
We have to mention here, that we are applying \textit{episodic} reinforcement learning, like it is typically performed for learning games. This means that multiple steps as described in Section~\ref{ssec:learner} form a so called \textit{episode}, in which we transition from an \textit{initial} configuration to a \textit{final} configuration. For games, the initial configuration typically describes the starting setup of the game board and the final configuration a state of winning or losing. In our problem of index selection, the initial configuration resembles having no indexes at all, whereas a final configuration is a configuration of the maximum number of indexes~$k$. Consequently, we form a new episode by, starting with the initial configuration, incrementally adding indexes one by one until the maximum amount~$k$ is reached. As soon as $k$ is reached, we conclude the episode, drop all indexes, and start the next episode with the initial configuration again. Thus, an action \texttt{drop\_index\_on($C_j$)} is \textit{not} required in this setup.

\subsection{Reward}
\label{ssec:reward}
Defining the reward function is a delicate task and has a high impact on the learning quality. In our example, the overall goal is to optimise the query response time. First, we have to define a cost function. If $L$ is the set of indexed columns, then $cost(L)$ returns the cost of the entire workload for this set of indexes. Of course, this cost can be either estimated by the DBMS to achieve a high training speed (e.g. by using the \texttt{EXPLAIN} command) or it can originate from actual query runtimes to achieve higher precision. 

To have a baseline to compare with, we first compute the cost of executing the workload without any indexes as ~$cost({\emptyset})$. Then, for an index configuration~$L$ containing $l$ indexes with~$l\in{1...k}$, the function $$r(L)=max\bigg(\frac{cost(\emptyset)}{cost(L)}-1,0\bigg)$$ returns the reward of that configuration~$L$. Obviously, the lower the~$cost(L)$, the higher the reward. If a configuration~$L$ does not lower the cost at all or even increases the cost, the reward is~$0$.

\subsection{Hyper Parameters}
\label{ssec:parameters}
To run the described reinforcement learner, a number of hyper parameters must be set before execution. 
 First, we have to set the upper limit on the amount of indexes, denoted as $k$, usually dictated by a space constraint on the available memory. We chose~$k=3$ for the upcoming evaluation. The remaining parameters are directly connected to the reinforcement learner and are largely chosen in a process of trial and error: we set the number of hidden layers to~$4$, the number of neurons per hidden layer to~$8$, and the activation function for the hidden layers and the output layer to \texttt{RELU} and \texttt{SOFTMAX} respectively. As agent, which resembles the actual learning algorithm, we use \texttt{CEM}.

\subsection{Design Advisory Tools}
As mentioned, there is a classical helper tool of the DBA with a similar purpose --- the design advisory software, that is shipped with an industry-grade DBMS. These tools are used by passing a workload file containing a set of SQL queries. To come up with index recommendations, they introduce so called \textit{virtual indexes}. A virtual index is just the description of a hypothetical index in form of meta-data, that is written into the catalog. No actual physical index structure is built. By this, the query optimiser is tricked in believing that a variety of indexes exist and considers them in finding the best query plan. When the best query plan is found, a recommendation for index creation is returned for every column on which an index access occurs in that plan. As a consequence of this design, the recommendation of the advisory tool is only as good as the statistics of the query optimiser. Besides, these tools typically compute a recommendation per query and then try to synthesise a workload recommendation from the individual results. Thus, the overall best decision on index selection might be missed.

\section{Evaluation}
\label{sec:evaluation}
With the description of the learning workflow and the problem environment at hand, let us come to the actual evaluation. In the following, we will compare the quality of index selection using deep reinforcement learning on a number of given workloads.    

\subsection{Setup}

All experiments are conducted on a desktop PC, equipped with a $4.5$~GHz Intel Core~i7 7700K processor with $4$~cores, 16GB of RAM, and a NVIDIA GeForce GTX~1080~Ti. For learning, we use \textit{keras-rl}~\cite{kerasrl}, that implements a general reinforcement learner in python on top of \textit{keras}, which is itself a high-level neural network API on top of a deep learning backend like \textit{Tensorflow}, \textit{Theano}, or \textit{Microsoft CNTK}. We chose Microsoft Cognitive Toolkit (CNTK) with GPU support as a backend to keras in this work.
To describe the problem environment, we use the \textit{gym}~\cite{gym} library from the OpenAI~\cite{openai} project and then pass this environment to keras-rl to perform the learning. For learning, we use the optimisation of \textit{experience replay}~\cite{human}, where the network is not directly trained on the current experience, but on previously seen ones to avoid getting stuck in local minima.
As database, we use schema and data of the TPC-H benchmark~\cite{tpch} in scale factor~$1$ and run the queries in PostgreSQL. In this work, our workloads query only the \texttt{LINEITEM}~table.

\subsection{Experiments}
\label{ssec:experiments}
For the experimental evaluation, we have to distinguish between the \textit{training workloads} and a \textit{test workload}. We use the training workloads to train the index selection during reinforcement learning. Note that we use a multitude of randomly generated training workloads to confront the learner with a variety of different scenarios. The test workload is then used to compare the quality of our trained neural network in comparison to having all indexes available.

\begin{table}[!htb]
	\setlength{\tabcolsep}{3pt}
	\begin{center}
		\begin{tabular}{| L{3cm} || R{4cm} | R{4cm} | R{4cm} |}
			\hline
			\textbf{Workload} & \textbf{NoIndex} [ms] & \textbf{IndexedAll} [ms] & \textbf{NoDBA} [ms]\\
			\hline\hline
			% 50 cols, small pages
			%\hline
			$W_1: Q1$ & $562.143$ & \color{blue}$13.912$ & $38.350$ \\\hline
			$W_1: Q2$ & $530.348$ & \color{blue}$31.868$ & $38.921$ \\\hline
			$W_1: Q3$ & $490.926$ & $82.439$ & \color{blue}$74.666$\\\hline
			$W_1: Q4$ & $552.060$ & $46.091$ & \color{blue}$37.583$\\\hline
			$W_1: Q5$ & $545.870$ & $39.811$ & \color{blue}$24.067$ \\\hline
			$W_1:$~Total & $2681.347$ & $214.121$ & \color{blue}$213.587$ \\\hline
			\hline
			$W_2: Q1$ & $498.074$ & \color{blue}$127.356$ & $276.205$ \\\hline
			$W_2: Q2$ & $491.517$ & $999.767$ & \color{blue}$193.690$ \\\hline
			$W_2: Q3$ & $548.675$ & $29.630$ & \color{blue}$24.653$ \\\hline
			$W_2: Q4$ & $548.188$ & $25.126$ & \color{blue}$24.617$ \\\hline
			$W_2: Q5$ & $543.516$ & $25.123$ & \color{blue}$24.086$ \\\hline
			$W_2:$~Total & $2629.97$ & $1207.002$ & \color{blue}$543.251$ \\\hline
			\hline
			$W_3: Q1$ & $705.850$ &  \color{blue}$0.057$ & $1028.729$ \\\hline
			$W_3: Q2$ & $737.364$ &  \color{blue}$1.375$ & $977.043$  \\\hline
			$W_3: Q3$ & $728.996$ & $803.244$ &  \color{blue}$743.211$  \\\hline
			$W_3: Q4$ & $644.503$ &  \color{blue}$0.923$ & $621.37$  \\\hline
			$W_3: Q5$ & $851.598$ & $1065.513$ &  \color{blue}$6.602$  \\\hline
			$W_3:$~Total & $3668.311$ &  \color{blue}$1871.112$ & $3376.955$ \\\hline
%			Number of episodes with the same workload & ? \\	
			\hline
		\end{tabular}
		\caption{Comparison of \textit{workload execution times}. We compare \textit{NoIndex} (executing the workload without any indexes) with our \textit{NoDBA} (executing the workload with the predicted indexes). Additionally, we show \textit{IndexedAll}, where an index on every column is available.}
		\label{table:performance}
	\end{center}
\end{table}

Every query in the training workloads is a \texttt{SELECT count(*)} query on the \texttt{LINEITEM}~table of TPC-H. In the \texttt{WHERE}~clause, we filter on a fixed number of columns and combine their individual selections using \texttt{AND}. For every randomly chosen column in the \texttt{WHERE} clause, we perform either an equality selection (e.g. \texttt{L\_TAX = 0.02}) or a range selection (e.g. \texttt{L\_SHIPDATE < '1994-01-01'}). The predicates are randomly selected from a list of actually occurring values. %Additionally, we allow the generation of multiple selections on a single column, which are combined by~\texttt{OR} (e.g. \texttt{L\_TAX = 0.01 OR L\_TAX = 0.02}), to realise lower selectivities on a certain column as well.

\begin{table}[!htb]
	\setlength{\tabcolsep}{3pt}
	\scriptsize
	\begin{center}
		\begin{tabular}{| C{1cm} || L{14.7cm} |}
			\hline
			\textbf{Query} & \textbf{Selections}\\
			\hline
			%\hline
			$Q1$ & \texttt{l\_partkey < 100, l\_suppkey < 100, l\_linenumber = 1, l\_discount = 0.02}  \\\hline	
			$Q2$ & \texttt{l\_orderkey < 100000, l\_partkey < 10000, l\_quantity = 1, l\_linenumber = 1}  \\\hline
			$Q3$ & \texttt{l\_quantity = 1, l\_partkey < 100000, l\_suppkey < 1000, l\_orderkey < 100000}  \\\hline		
			$Q4$ & \texttt{l\_orderkey < 100000, l\_discount = 0.0, l\_suppkey < 1000, l\_linenumber = 1}  \\\hline
			$Q5$ & \texttt{l\_orderkey < 100000, l\_partkey < 100000, l\_suppkey < 10000, l\_linenumber = 1, l\_discount = 0.01}  \\\hline				
		\end{tabular}
		\caption{Workload~$W_1$}
		\label{table:workload1}
	\end{center}
	\vspace{-0.5cm}
\end{table}

\begin{table}[!htb]
	\setlength{\tabcolsep}{3pt}
	\scriptsize
	\begin{center}
		\begin{tabular}{| C{1cm} || L{14.7cm} |}
			\hline
			\textbf{Query} & \textbf{Selections}\\
			\hline
			%\hline
			$Q1$ & \texttt{l\_orderkey < 1000000, l\_partkey < 10000, l\_suppkey < 10000, l\_linenumber = 1, l\_quantity = 1}  \\\hline	
			$Q2$ & \texttt{l\_quantity = 1, l\_partkey < 100000, l\_orderkey < 1000000, l\_linenumber = 1}  \\\hline
			$Q3$ & \texttt{l\_orderkey <
100000, l\_partkey < 100000, l\_suppkey < 100000, l\_quantity = 1, l\_discount = 0.0}  \\\hline		
			$Q4$ & \texttt{l\_orderkey < 100000, l\_partkey < 100000, l\_suppkey < 100000, l\_linenumber = 1, l\_quantity = 1, l\_discount = 0.02}  \\\hline
			$Q5$ & \texttt{l\_orderkey < 100000, l\_partkey < 100000, l\_suppkey < 100000, l\_linenumber = 1, l\_discount = 0.02}  \\\hline					
		\end{tabular}
		\caption{Workload~$W_2$}
		\label{table:workload2}
	\end{center}
	\vspace{-0.5cm}
\end{table}

\begin{table}[!htb]
	\setlength{\tabcolsep}{3pt}
	\scriptsize
	\begin{center}
		\begin{tabular}{| C{1cm} || L{14.7cm} |}
			\hline
			\textbf{Query} & \textbf{Selections}\\
			\hline
			%\hline
			$Q1$ & \texttt{l\_orderkey < 10, l\_suppkey < 50000, l\_extendedprice < 50000, l\_receiptdate < '1993-12-31', l\_returnflag = 'A', l\_linestatus = 'O'}  \\\hline	
			$Q2$ & \texttt{l\_orderkey < 1500, l\_extendedprice < 10000, l\_shipinstruct = 'TAKE BACK RETURN', l\_receiptdate < '1993-06-30'}  \\\hline
			$Q3$ & \texttt{l\_suppkey < 100, l\_shipdate < '1993-01-01', l\_receiptdate < '1992-06-29', l\_linenumber = 4, l\_shipinstruct = 'TAKE BACK RETURN', l\_shipmode = 'SHIP'}  \\\hline		
			$Q4$ & \texttt{l\_orderkey < 1500, l\_suppkey < 1000, l\_shipdate < '1995-03-31', l\_linenumber = 4, l\_tax = 0.0, l\_returnflag = 'N'}  \\\hline
			$Q5$ & \texttt{l\_suppkey < 50000, l\_extendedprice < 1000, l\_commitdate < '1995-01-28', l\_receiptdate < '1992-06-29', l\_quantity = 1, l\_linestatus = 'O'}  \\\hline					
		\end{tabular}
		\caption{Workload~$W_3$}
		\label{table:workload3}
	\end{center}
\end{table}

For testing the quality, we use three synthetic workloads, each consisting of five queries of the form \texttt{SELECT COUNT(*) FROM LINEITEM WHERE Selection1 AND Selection 2 AND ...}, where the selections are listed in Tables~\ref{table:workload1}, \ref{table:workload2}, and \ref{table:workload3}.
The workloads~$W_1$ (Table~\ref{table:workload1}) and $W_2$ (Table~\ref{table:workload2}) consist of queries, which perform up to six selections with high selectivity on a fixed set of attributes of~\texttt{LINEITEM}. The queries of workload~$W_3$ (Table~\ref{table:workload3}) choose their selections on all attributes of \texttt{LINEITEM}. In Table~\ref{table:performance}, we now evaluate the performance of running the workload without indexes as well as indexes on all columns in comparison with the recommended ones by our NoDBA. As we can see, on the highly selective workloads~$W_1$ and $W_2$, our selection performs as good or even better\footnote{The query optimiser might make a bad decision in using an index on a particular column.} as having indexes on all columns. For~$W_3$, which selects on a larger set of attributes, our selection of only~$3$ indexes still manages to improve the runtime. The training time of our network, that has to happen once, was $42$~minutes. After that, an individual prediction took only around~$20$ms.

\section{Conclusion and Future Work}

In this work, we demonstrated how deep reinforcement learning can be used to automatically administer a database management system. We showcased our concept using index selection as \textit{one} possible example. We showed that it is possible to train a neural network for such a task.

	Obviously our work is just a first glimpse on what is possible when tuning a database using deep learning. We are currently working on extending these initial results in several dimensions.

	For instance, compared to traditional cost-based what-if optimisations, a strong benefit of our approach is that we can also optimise \textit{without} those estimates --- which are possibly faulty and misleading anyways~\cite{LeisGMBK015}. We are currently exploring the trade-offs of this idea.

	Motivated by these promising results, we also believe that deep reinforcement learning can be used to tune a variety of other components of a DBMS, such as the query optimiser. In addition, we would like to extend our work to non-stationary workloads.

\bibliographystyle{abbrv}
\bibliography{bibliography}

\end{document}